\documentclass[twocolumn,aps,prb,unsortedaddress,superscriptaddress,showpacs]{revtex4}

\input epsf

\usepackage{graphicx}
\usepackage{dcolumn}
\usepackage{bm}
\usepackage{color}

\begin{document}

\preprint{Kienle: EHT-CNT}

\title{Extended H\"uckel theory for bandstructure, chemistry and transport.\\ I. Carbon Nanotubes}
\author{D. Kienle}
\email{kienle@ecn.purdue.edu}
\affiliation{Purdue University, Department of Electrical and Computer Engineering, West Lafayette, IN 47907, USA}
\author{J.I. Cerda}
\affiliation{Instituto de Ciencia de Materiales de Madrid, CSIC, Cantoblanco 28049, Madrid, Spain}
\author{A.W. Ghosh}
\affiliation{University of Virginia, Department of Electrical and Computer Engineering, Charlottesville, VA 22903, USA}

\date{\today}

\widetext
\begin{abstract}
We describe a semi-empirical atomic basis Extended H\"uckel Theoretical (EHT) technique 
that can be used to calculate bulk bandstructure, surface density of states, 
electronic transmission and interfacial chemistry of various materials within the same
computational platform. We apply this method to study multiple 
technologically important systems, starting with carbon-nanotubes (CNT) and their interfaces
in this paper, and silicon-based heterostructures in our follow-up paper. 
We find that when it comes to quantum transport through interesting, complex heterostructures,
the H\"uckel bandstructure offers a fair and practical compromise between orthogonal tight-binding 
theories (OTB) with limited transferability between environments under large 
distortion, and density functional theories (DFT) that are computationally quite expensive
for the same purpose.
\end{abstract}

\pacs{61.46.Fg, 73.43.Cd, 73.63.Fg}


\maketitle

\section{Introduction}
Quantitative electronic structure theories are essential to the
understanding and designing of novel materials and devices. It is now
generally accepted that transport properties of nanoscale devices
depend on both the intrinsic electronic structure of the active
channel, as well as its interfacial properties with contacts and other
scattering centers. A particular challenge in this respect is to
incorporate both long and short range correlations within the same
framework, such as the bulk bandstructure of periodic solids as well as
the local chemical properties of clusters, surfaces, and interfaces.
For instance, simulating scanning tunneling spectra (STS) of molecules
on silicon substrates requires an accurate description of the silicon
bulk bandstructure that quantitatively captures not just the bandgap
responsible for the onset of negative differential resistance,
\cite{RakshitNDR} but also the multiple effective masses which determine
the contact density of states and injection velocities, and the strain
parametrizations that capture atomic reconstruction and relaxation near
the surface and their bonding with molecular components.\cite{LiangC60} 
In addition, one needs to describe the electrostatics
responsible for band-bending in the silicon depletion layer, the
molecular transport levels and their transmission under bias, and
finally the electronic properties of the scanning tip and the
intervening vacuum layer, all within the same formalism. It is no
surprise therefore that under these circumstances, standard electronic
structure techniques developed by quantum chemists for simulating
molecules are usually incompatible with those developed by solid-state
physicists for bulk bandstructure, making it important to develop a
common formalism that addresses both domains of interest and also
maintains a good compromise between computational accuracy and
practicality.

While sophisticated methods exist for equilibrium geometry and
bandstructure, comparable success has yet to be achieved for transport
problems, partly because of the lack of universally accepted
experimental standards, but mainly because quantum transport inherently
involves solving a complicated nonequilibrium open boundary problem for
which electronic structure theories are not benchmarked. A proper
quantitative understanding of correlation effects in transport is still
evolving and it is not yet clear if mean-field approaches that work at 
equilibrium are at all capable of handling the profusion of electronic 
excitations that often dominate in nanoscale 
conduction.\cite{McEuenKondo,MuralidharanFockSpace}
Even aside from such correlation issues, one
needs to worry about heterointerfaces since current flow involves
charge transport across two intrinsically different material systems --
a multimoded contact consisting of a highly conductive material
externally maintained at thermal equilibrium, and a sparsely moded
device region that is readily driven away from equilibrium and acts as
the active transport channel.

In this paper, we employ a semi-empirical approach to electronic
structure that can be adopted for electronic conduction through
complex hybrid systems by combining it with the Nonequilibrium Green's
function (NEGF) technique for quantum transport. Our theoretical
parameters are tailored to salient features of the bulk bandstructure,
while the employment of non-orthogonal basis sets resembling linear
combinations of underlying atomic orbitals seems to make them fairly
transferrable to surfaces as well, as observed in the
past.\cite{GoringeRepTB,PecciaTrans} In addition, the presence of
explicit basis sets opens up the possibility of `stitching' together
disparate regions\cite{KienleMetalCNT} by matching the interfacial
Green's function, which is the {\it{only}} quantity through which the
diverse regions communicate with each other quantum mechanically. The
modularity of our approach also allows us to conveniently separate the
problems of determining the optimized interfacial geometry and the
interfacial transmission (we are ignoring current-induced forces), the
former depending on the total energy of the system while the latter
depends only on a few relevant single-electron levels near the Fermi
energy. In other words, {\it{given a particular atomistic configuration
of the contact-channel-contact heterostructure, we seek to determine
its transport properties by coupling our electronic structure approach
with quantum transport using NEGF.}}

The outline of the paper is as follows: section II explains the
strengths of EHT over other traditional bandstructure methods. 
In section III we briefly summarize the NEGF aproach used to calculate 
density of states and transmission of CNTs; we then specify the model 
Hamiltonian and describe the details of EHT used to determine the bandstructure. 
The numerically calculated bandstructure data for nanotubes are then compared 
in section IV with experimental scanning tunneling spectroscopy experiments 
along with other theoretical approaches. In section V we investigate the changes
in the dispersion of a semi-conducting CNT under large lateral
deformations as well as with a CO molecular attachment to its surface
that allows it to function as a molecular sensor. We summarize our work
and discuss future extensions in the last section.

\section{Why this particular method?}

A particular trade-off in any bandstructure theory is between
flexibility and rigor. While empirical, orthogonal tight-binding (OTB)
methods are quick and practical, they are benchmarked for specific
geometries and are usually not very transferable to other environments
involving significant structural deformations beyond a few percent.
Tight binding basis-sets are commonly assumed to be both orthogonal and
short-ranged,\cite{GoringeRepTB} while atomic wavefunctions are not,
meaning that OTB basis sets do not resemble eigenstates of an atomic
Hamiltonian. Efforts at improving tight-binding theories involve going
beyond nearest neighbor techniques, using higher virtual orbital bases for
increased completeness,\cite{VoglSP3S*,Jancu1_SP3D5S*,Boykin_TBsp3ds*}
and employing power laws for parameter transfer under small ($\sim 1-2 \%$)
strain.\cite{KeatingStrain} Nevertheless these models are likely
to miss important chemical details involving properties varying on
an interatomic length scale, in particular near deformed surfaces and
interfaces where a drastic reconstruction of the atomic structure is
expected.\cite{GoringeRepTB}

At the other end of the spectrum are accurate, but computationally
expensive first-principles techniques developed by quantum chemists and
solid state physicists, such as Configuration Interaction (CI) and
Density Functional Theories (DFT) in various atomic or plane wave basis
sets or combinations thereof. Structural deformations are naturally
captured by such total energy calculations by solving a one electron
Schr\"odinger equation in a suitable self-consistent potential
approximating the electron-electron
interaction.\cite{ParrDFT,BarthDFT,MartinElStruct} Such codes are
typically based on rigorous variational theorems and are quantitatively
quite accurate, at least for equilibrium properties.  Their extension
and practical implementation to transport beyond the linear response
regime is continuously evolving,\cite{GuoDFT,DamleDFT,StokbroDFTNEGF}
and a topic of current research.\cite{DiVentraDFTLDA} Conceptually, it
is not clear if any self-consistent potential approach can quantitatively 
describe the rich spectrum of many-body transitions that are often experimentally 
accessed in strongly correlated transport in weakly coupled 
systems.\cite{McEuenKondo,MuralidharanFockSpace}

We aim for a practical compromise between these two limits by using a
semi-empirical technique motivated by Extended-H\"uckel calculations
popular in the chemistry community.  Such theories, widely used in the
past to describe the equilibrium properties of isolated
molecules,\cite{Murrell1972} have recently been applied to molecular
conduction \cite{Tian} and also extended to solids using transferable
atomic-orbital basis sets (AO) for calculating the electronic structure
of various compounds benchmarked with detailed DFT calculations within
the local density (LDA) or generalized gradient (GGA) approximations.\cite{CerdaEHT}
Given a geometry, one uses the explicit EHT basis
functions to calculate a non-orthogonal overlap matrix $S$, which along
with separately fitted onsite Hamiltonian matrix elements yields the
corresponding off-diagonal hopping elements of the Hamiltonian. Within the
standard H\"uckel prescription, structural changes are simply accounted
for by re-calculating the overlap and hopping elements, but leaving the
basis sets and onsite elements unchanged.

In the following, we apply this EHT parametrization scheme\cite{CerdaEHT}
by benchmarking it to a two-dimensional graphene sheet\cite{CerdaWEB,SolerSiesta} 
and extending it to obtain the bandstructure, density of states and electronic
transmission of carbon nanotubes (CNT) of varying chiralities. We
show that the same bulk-optimized EHT-parameters (onsite energies and
AO-basis functions) are transferable to small diameter CNT bandstructures,
capturing even curvature-induced bandgap effects for larger than $1-3 \%$
tube deformation, in quantitative agreement with STS data.  Furthermore,
surface chemical effects are examined through the study of nanotube
based carbon-monoxide sensors whose alteration of electronic structure
upon molecular adsorption compares favorably with ab-initio calculations
of da Silva {\it{et al.}}.\cite{DaSilvaCO} In our follow-up paper, we
will demonstrate a similar transferability between bulk silicon, various
silicon surfaces, apply the EHT-methodology to unreconstructed silicon nanowires.
Taken together, the wide variety of these examples illustrates the range of 
transferability of EHT parameters, which we believe makes Extended H\"uckel 
Theory a useful practical tool for electronic structure and quantum transport.

\section{Approach}

Simulating conduction through a heterostructure involves combining
suitable bandstructures for the channel and contact materials with
self-consistent electrostatics and quantum transport.  While the
formulation of correlated transport is itself an active area of research,
our aim here is to develop a minimal model that would capture quantum
chemistry, surface physics, bandstructure and electrostatic effects
that are all crucial for different prominent aspects of nanoscale
conduction. The ingredients needed for a proper simulation are the
Hamiltonian and potential matrices describing the device bandstructure
and electrostatics, and the contact self-energies that effectively open
up the system and allow us to add or remove charge under nonequilibrium
conditions. The EHT prescription gives us a practical way to calculate
these ingredient matrices for a given atomistic structure, and then
connect them with a non-equilibrium Green's function
(NEGF) formulation of quantum transport, which we briefly summarize
below.\cite{HaugKinetic1992,DattaMeso1995} The retarded Green function
of the device is given by 
\begin{equation} 
G = \left[ (E+i\eta) S - H - \Sigma_l - \Sigma_r \right]^{-1}~, 
\end{equation} 
where $S$ and $H$
describe the overlap and Hamiltonian matrices of the device unit cell
calculated according to the H\"uckel prescription Eq.(\ref{EHT_Rule}).
The matrix elements $\Sigma_{l,r}$ are self-energies that provide open
boundary conditions to the device with the left and right semi-infinite 
contacts. The self-energy $\Sigma=\tau g \tau^{\dagger}$ incorporates the 
coupling matrix $\tau$ describing the contact-device bonding, while $g$ is 
the surface Green function of the left/right contact calculated by means of
a recursion technique.\cite{RubioTransferMatrix, RubioSurfGreen} In a
non-orthogonal tight-binding scheme the density of states (DOS) is given
by $D(E)=\frac{i}{2\pi} Tr\left( A S\right)$ where $A=i(G-G^{\dagger})$
denotes the spectral function. Finally, in the phase-coherent limit
the zero-bias transmission through the unit cell reads $T(E)=Tr\left[
\Gamma_l G \Gamma_r G^{\dagger} \right]$, where $\Gamma_{l,r}=i\left(
\Sigma_{l,r} - \Sigma_{l,r} \right)$ are the broadening matrices which
specify the time an electron resides within the device. In this paper, we
will study infinite nanotubes so that the active device is just one CNT
unit cell and the left and right contacts extend that cell to infinity
in either direction.

\begin{figure}[htbp]
\centerline{\epsfysize=2.75in
\epsffile{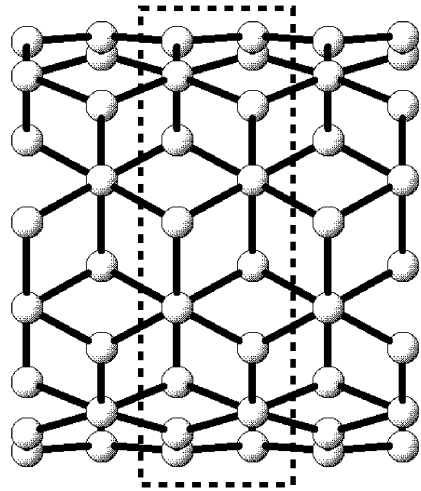}} 
\vspace{-0.3cm}
\caption{Sketch of a $(5,5)$ armchair tube with three unit cells.
The dashed rectangle marks the center unit cell containing $20$
carbon atoms. For the calculation of the $E-k$ dispersion we use
the 1st nearest-neighbor unit cells to the left and right,
respectively.} 
\label{Fig_3DTubeStructure}
\end{figure}

The bandstructure of a nanotube with chirality $(n,m)$ is calculated
employing the non-orthogonal Slater-Koster scheme and solving for the
generalized eigenvalue problem\cite{KosterNTB}
\begin{eqnarray}
{\bf H} ({\bf k}) A_i ({\bf k}) = E_i ({\bf k}) {\bf S} ({\bf k}) A_i ({\bf k})~,~
\end{eqnarray}
where $A_i ({\bf k})$ denotes the eigenvector of the $i^{th}$
subband, and ${\bf k}$ is a Bloch wavevector within the first 
Brillouin zone. The size of the overlap and Hamiltonian matrices
are determined by the chosen basis set, i.e., the number of atoms
within the unit cell multiplied by the number of orbitals per
atom. In our case, using four $sp$-orbitals per atom, the size of
these matrices is $80 \times 80$ for an armchair $(5,5)$ tube as
sketched in Fig.~\ref{Fig_3DTubeStructure}.
The overlap- and Hamilton matrices ${\bf S} ({\bf k})$ and 
${\bf H} ({\bf k})$ representing the structure in reciprocal space 
are calculated by
\begin{eqnarray} \label{SKHK}
{\bf H}_{i,j} ({\bf k}) &=& \sum_{j^\prime,m^\prime} e^{i {\bf k} \cdot \left( {\bf R}_{i0} - {\bf R}_{
j^\prime m^\prime}\right)}~{\bf H}_{i0,j^\prime m^\prime} \\
{\bf S}_{i,j} ({\bf k}) &=& \sum_{j^\prime,m^\prime} e^{i {\bf k} \cdot \left( {\bf R}_{i0} - {\bf R}_{
j^\prime m^\prime}\right)}~{\bf S}_{i0,j^\prime m^\prime}~,~
\end{eqnarray}
where $i,j$ label atoms within the unit cell, and $m^\prime$ is the unit cell
index. The summation indices in ${\bf H}_{i,j} ({\bf k}), {\bf S}_{i,j}
({\bf k})$ run over all atoms $j^\prime$ in unit cell $m^\prime$ which are
equivalent to atom $j$ in the reference unit cell $m=0$.  The real-space matrix
elements ${\bf H}_{i0,j^\prime m^\prime}~,~ {\bf S}_{i0,j^\prime m^\prime}$, between
an atom $i$ in the reference unit cell and atom $j^\prime$ in cell
$m^\prime$ are calculated by means of the Extended H\"uckel prescription.
%

Perhaps the most prominent difference between the empirical tight-binding and EHT
principle is the use of an irreducible representation of the tight-binding
Hamiltonian which is directly fitted to available bandstructure
data without employing explicit basis sets. In EHT however one works directly 
with the orbital basis functions out of which
the Hamiltonian elements are constructed using the H\"uckel principle. The
diagonal elements are benchmarked with experimental values of electronic
`hardness', i.e. the difference between ionization potential and
electron affinity.\cite{Murrell1972} The off-diagonal matrix elements
are then determined directly from the prescription
\begin{eqnarray} \label{EHT_Rule}
H_{\mu\mu} &=& E_{\mu\mu}, \nonumber\\
H_{\mu\nu} &=& \frac{1}{2} K_{eht}~S_{\mu\nu} \left( H_{\mu\mu} 
+ H_{\nu\nu}\right)~, \nonumber\\
S_{\mu\nu} &=& \int d^3 {\bf r}~\phi_{\mu}^{*} ({\bf r})~\phi_{\nu} ({\bf r})~,
\end{eqnarray}
where $\mu,\nu$ label the atomic orbitals, and $S_{\mu\nu}$ is the
overlap matrix between the orbital basis function $\phi_{\mu}$ and
$\phi_{\nu}$, respectively. $K_{eht}$ is an additional fitting
parameter with a value of $1.75$ commonly used for molecules and $2.3$
for solids.\cite{Murrell1972,CerdaEHT} In the case of the planar
2D-graphene sheet a good match is achieved for a value of $K_{eht} =
2.8$.\cite{CerdaWEB} One important assumption within EHT is that the
hopping matrix elements $H_{\mu\nu}~,~\nu\neq \mu$ depend linearly on
the overlap matrix $S_{\mu\nu}$ alone.\cite{Murrell1972} EHT-basis
functions are usually Slater-type orbitals (STO), $\Phi_{nlm} = N
r^{n-1}~e^{-\zeta r}~Y_{lm}(\Theta,\varphi)$, which have the correct
asymptotic form at large distance $r$ ($n$, $l$, and $m$ denote respectively
the principle, azimuthal and magnetic quantum numbers). The individual
orbital wavefunctions $|q\rangle$ are then approximated by a linear
combination of STOs, with coefficients and exponents $\left\{ c_i,
\zeta_i \right \}$ fitted for the individual basis functions to match
bandstructure data.  Since the basis sets are directly fitted to
experimental or theoretical data, the resulting AOs are more localized
than the free atomic wavefunctions, although they may still be regarded
as long-ranged compared to the usual TB Wannier-like description; we
typically use a cut-off interatomic distance of $R_c=9$ \AA~for the
interactions. The use of basis function that are not too localized
turns out to be a key issue for achieving a good
transferability.\cite{CerdaEHT} Admittedly, the use of a direction
independent $K_{eht}$ function is a drastic assumption that can be
relaxed by making the constant orbital dependent, but we make the
simplifying assumption that the orientation dependence arises
principally from the corresponding overlap functionals.

The H\"uckel prescription Eq.~(\ref{EHT_Rule}) can be generalized to
connect different chemical sub-systems $A, B$. The problem is that each
sub-system has its own parametrization that yields its own vacuum level
relative to which electronic levels are calculated. For instance, the
Fermi level of most metals is set to $E_F = -10$ eV in the paper in,\cite{CerdaEHT} 
which means that each dispersion curve needs to be
individually shifted to bring its Fermi level upto the experimental
value, through the transformation ${\bf H} \rightarrow {\bf H} + V_c
{\bf S}$ for each sub-system. The correct alignment of the levels
relative to each other becomes particularly important when studying
compound systems such as metal-semiconductor junctions or molecular
components attached on nanowires and nanotubes. We calculate the
coupling matrix across such a hybrid junction between subsystems A and
B using the H\"uckel principle Eq.(\ref{EHT_Rule}) through the interpolation scheme
\begin{eqnarray} \label{EHT_RuleModified}
\tilde{{\bf H}}_{\mu_A \nu_B} &=& \frac{1}{2}~{\bf S}_{\mu_A \nu_B}
\left[ \left( K_A {\bf H}^{A}_{\mu \mu} + V_{cA} \right) \right. \nonumber\\
&& \left. +
\left( K_B {\bf H}^{B}_{\nu \nu} + V_{cB} \right) \right]~~,
\end{eqnarray}
where $V_{cA}$ and $V_{cB}$ are the shifts needed to align the vacuum
levels for sub-systems $A$ and $B$, respectively. It is worth noting
that this approach provides a simple interpolation scheme that
gives us the correct limiting values of the Hamiltonian for the two
individual subsystems. Further work, however, clearly needs to be done 
to calibrate this interpolation scheme to specific interfacial properties 
such as measured charge transfer doping, Schottky barrier heights, or perhaps
first principles calculations of interfacial dipoles or chemisorption
properties such as workfunction modification at surfaces.

\section{Results: Nanotube Bandstructures}

\begin{figure}[htbp]
\centerline{\epsfysize=4.0in
\epsffile{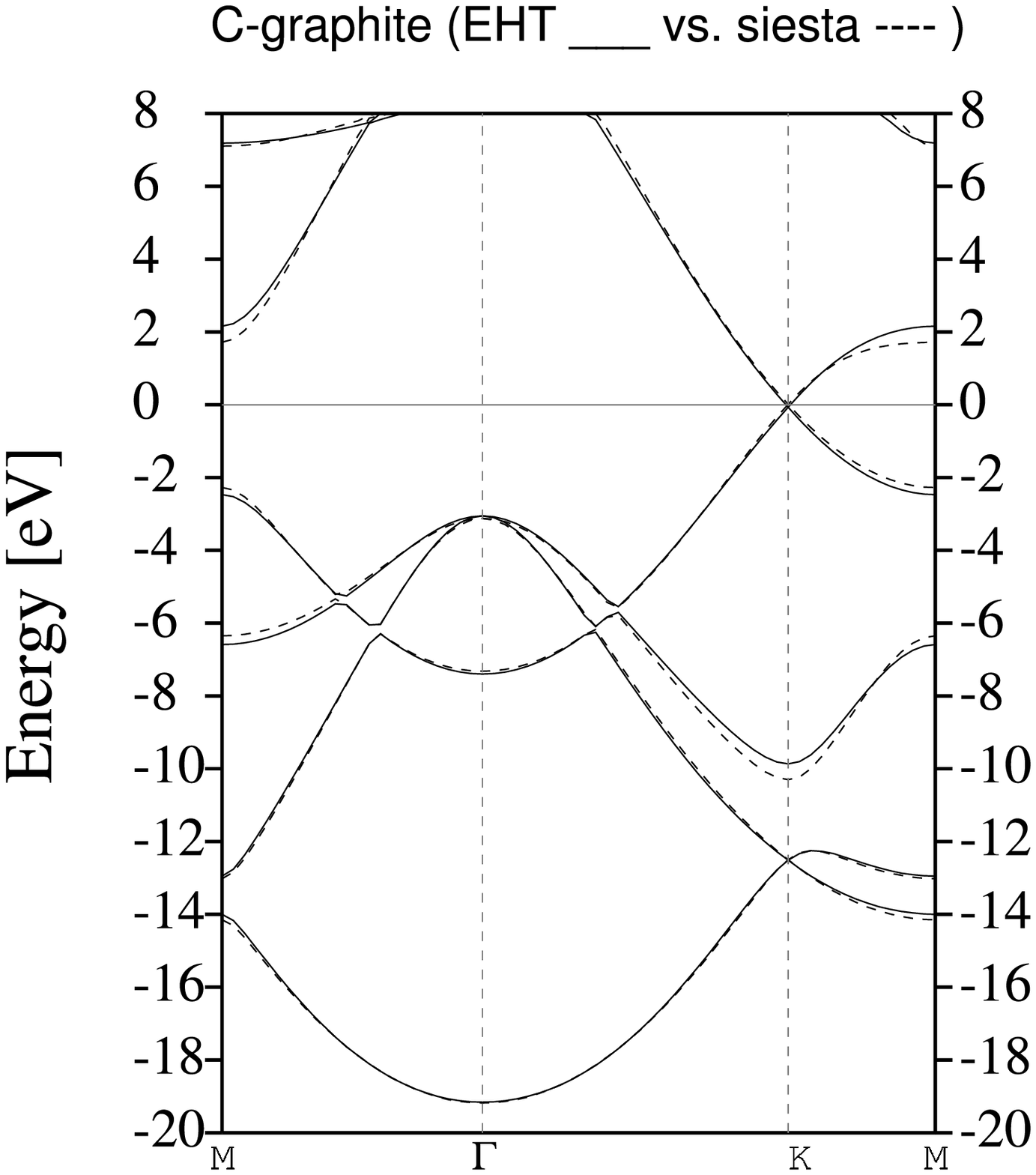}}
\vspace{-0.1cm}
\caption{2D bandstructure of a graphene sheet along the $M \rightarrow
\Gamma \rightarrow K \rightarrow M$ path within the 1st Brillouin
zone. The solid line corresponds to the EHT-bandstructure using
parameters optimized to match the DFT-GGA bandstructure (dashed
line).\cite{CerdaWEB,SolerSiesta} The C-C bonding distance is set to 
$a_{C-C} = 1.44$ \AA, and the cut-off radius for the neighbor atoms is 
$R_c = 9$ \AA. The Fermi level is at $E_F=0.0$ eV.}
\label{Fig_GrapheneEK_Cerda}
\end{figure}

We begin by benchmarking our parameters with a two-dimensional 
graphene sheet. Figure~\ref{Fig_GrapheneEK_Cerda} shows the corresponding
dispersion relation along the $M \rightarrow \Gamma \rightarrow K
\rightarrow M$ path within the 2D-Brillouin zone. The dashed line in
Fig.\ref{Fig_GrapheneEK_Cerda} is the DFT-GGA calculation of the $E-k$
dispersion calculated using the SIESTA code,\cite{CerdaWEB,SolerSiesta}
to which the EHT bandstructure is fitted (solid line) by
adjusting the onsite Hamiltonian matrix-elements $H_{ii}$,
the exponentials and the expansion coefficients of the Slater-orbital
basis functions. As atomic like basis functions for each carbon atom 
we are using two basis sets: (i) $sp^3$ and (ii) $sp^3 d^5$ orbitals, each of which 
has been optimized to match the DFT-GGA bandstructure. The two parameter sets are
given in Table \ref{TableEHTCarbon}.
In our calculation all atoms within a cut off radius of $9$ \AA~ from the two non-equivalent 
atoms within a unit cell are included. Furthermore, we set the Fermi level of graphene to 
$E_F=0.0$ eV and $K_{\rm{eht}}$ is set to 2.8.\footnote{Note, that in the EHT-parameter 
optimization the Fermi level for graphene is set to $E_F=-13$ eV, so that the bands are 
unique up to an overall shift.\cite{CerdaEHT,CerdaWEB} The role of the Fermi energy
in parameter fitting has been addressed in section III. Its absolute 
value, however, becomes particularly important for compound systems as shown in section V.}.
%
%

%
\begin{table}[htbp]
\begin{tabular}[t]{ c | c | c | c | c | c | c  }
\hline\hline
         &  AO    &  $E_{on}$   &  $\zeta_1$  &  $c_1$    &  $\zeta_2$  &  $c_2$    \\ \hline\hline
C-sp     &  $2s$  &  $-20.316$  &  $2.037$    &  $0.741$  &             &           \\
         &  $2p$  &  $-13.670$  &  $1.777$    &  $0.640$  &  $3.249$    &  $0.412$  \\ \hline\hline
C-spd    &  $2s$  &  $-19.889$  &  $2.025$    &  $0.764$  &             &           \\
         &  $2p$  &  $-13.080$  &  $1.624$    &  $0.272$  &  $2.177$    &  $0.739$  \\
         &  $3d$  &  $-2.046$   &  $1.194$    &  $0.491$  &             &           \\ 
\hline\hline
\end{tabular}
\caption{EHT parameters for carbon fitted to the 2D-bandstructure of graphene 
calculated using DFT-GGA.\cite{CerdaWEB,SolerSiesta} For both parameter sets the
parameter $K_{eht}$ was set to $2.8$.}
\label{TableEHTCarbon}
\end{table}

Carbon-nanotubes with chirality $(n,m)$ are obtained by wrapping the
two dimensional graphene sheet along specific circumference vectors
${\bf C}(n,m) = n {\bf a}_1 + m {\bf a}_2$.\cite{SaitoCNT} Note that
this approach, being truly atomistic, goes beyond the conventional
zone-folding scheme.  In the following, we will examine the
transferability of the two EHT parameter sets (sp and spd) of graphene to 
bandstructures for small diameter nanotubes.
In our calculation, we assume that structural variations of the CNT affect its hopping 
elements $H_{ij}$ only through the overlap matrix $S_{ij}$ (cf. Eq.~\ref{EHT_Rule}). 
This assumption means that the re-distribution of charge due to structural changes is
discarded, so that the bandstructure is determined in a non
self-consistent manner. For all tubes considered here we include the
coupling of nearest neighbor unit cells which consist of two rings each
in the case of armchair $(n,n)$ tubes (translation vector $T_0=2.39$ \AA), 
and four rings for zig-zag $(n,0)$ CNTs ($T_0=4.32$ \AA).

\subsection{Metallic armchair tubes}

\begin{figure}[htbp]
\centerline{\epsfysize=2.75in \epsffile{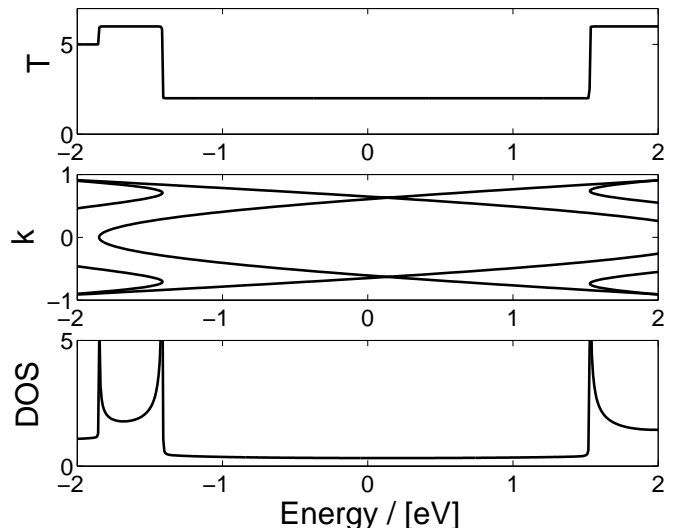} }
\vspace{-0.2cm}
\caption{Bandstructure, transmission and DOS per spin for an armchair $(5,5)$ tube 
using $sp^3$-orbitals. Due to their mirror symmetry the curvature of the tube does 
not break this symmetry, so that the tube remains strictly metallic. The Fermi level 
is $E_F=0.0$ eV and $k$ is in units of $[\pi/T_0]$.}
\label{Fig_Armchair}
\end{figure}

Fig.~\ref{Fig_Armchair} shows the bandstructure, transmission and
density-of-states (DOS) for a $(5,5)$ armchair tube within a $sp^3$-EHT model. 
The DOS shows typical features of a one-dimensional system, with a constant 
value within an energy interval of $[-1.5,1.5]$ eV, and van Hove singularities at
the onsets of higher subbands.  The transmission per spin within the
interval $[-1.5,1.5]$ eV is $2$ indicating that two bands can
in principle contribute to transport. Including spin, one should then
observe a maximum linear response (zero bias) conductance of $G = 2G_0$
with $G_0 = 2e^2/h$ assuming no parasitic resistances posed by contact
interfaces. Notably, curvature effects do not disrupt the bandstructure
of armchair tubes, which stay metallic because the mirror symmetry is
not broken 
upon wrapping the graphene sheet.
%

\subsection{Curvature effects on non-armchair `semi-metallic' tubes}

Experimentally it is known that small diameter carbon nanotubes that
are normally expected to be metallic by the $m-n$ rule,\cite{SaitoCNT}
with $(9,0)$ and $(12,0)$ chiralities for example, exhibit a
curvature-induced gap than $k_B T \approx 25$ meV\cite{LieberSTS} at
the Fermi energy. A simple $\pi$-orbital tight-binding
model,\cite{SaitoCNT,RocheMetSemMet,TaoCNTQD} commonly employed in CNT
transport simulations usually does not account for this effect, but
such s-p hybridizations might be important to include when considering
CNTs as possible candidates for interconnects. For tubes with
diameters $d \le 1$ nm, the structural deformation of the graphene
sheet affects its electronic structure significantly enough that such
an opening of a bandgap can be induced. The opening arises due to a
reduction of the overlap between the nearest-neighbor $\pi$-orbitals
under deformation causing the Fermi wavevector $k_F$ to move away from
the K-point within the 1st Brillouin zone.\cite{HanCNT,LieberSTS}
%
\begin{figure}[htbp] 
\centerline{\epsfysize=2.75in \epsffile{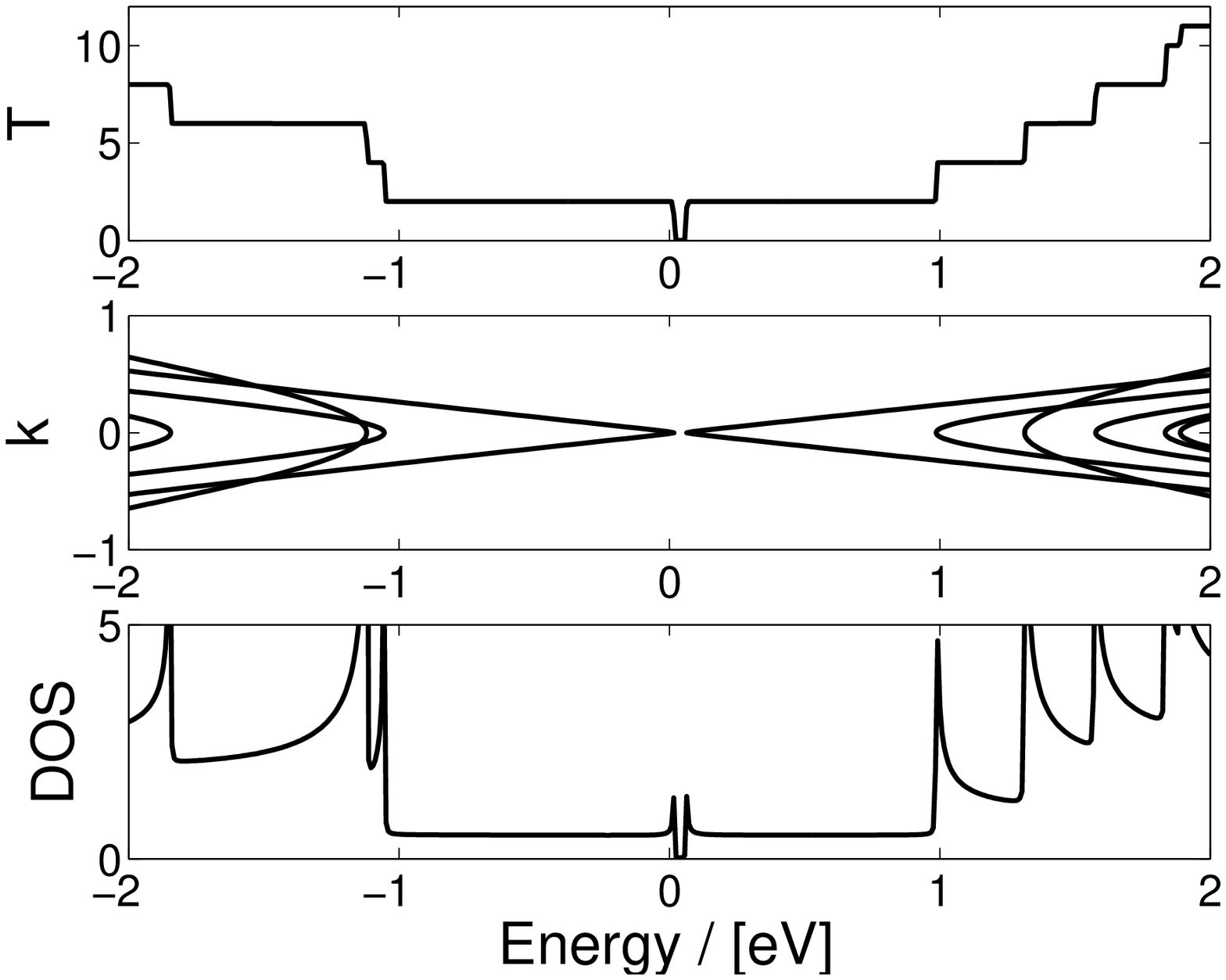}}
\vspace{0.1cm}
\centerline{\epsfxsize=2.6in \hspace{0.3cm}\epsffile{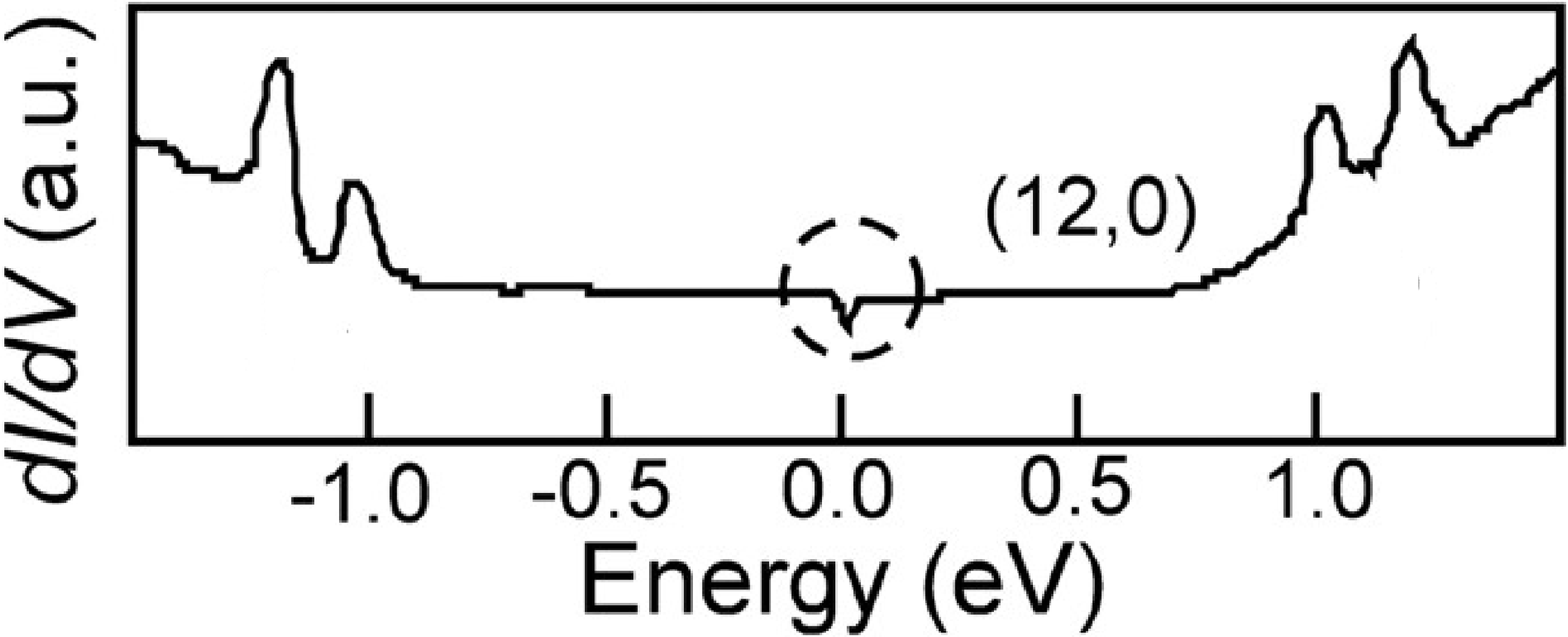}}
\vspace{-0.3cm}
\caption{Bandstructure, DOS, and transmission per spin for a ``metallic''
zig-zag $(12,0)$ tube using $sp^3$-orbitals. The gap close to the Fermi level at 
$E_F=0.0$ eV is about $50$ meV.  For comparison the experimental STS-$dI/dV$ signal
is shown at the bottom\cite{LieberSTS}. The bottom figure has been 
``Reprinted (abstracted/excerpted) with permission from 
M. Ouyang, J.L. Huang, C.L. Cheung, and C.M. Lieber, ``Energy Gaps in Metallic 
Single-Walled Carbon Nanotubes,'' {\em Science}, {\bf 292}, 702 (2001). 
Copyright 2001 AASS.''}
\label{Fig_N12M0} 
\end{figure}

\begin{figure}[htbp] 
\centerline{\epsfysize=2.75in \epsffile{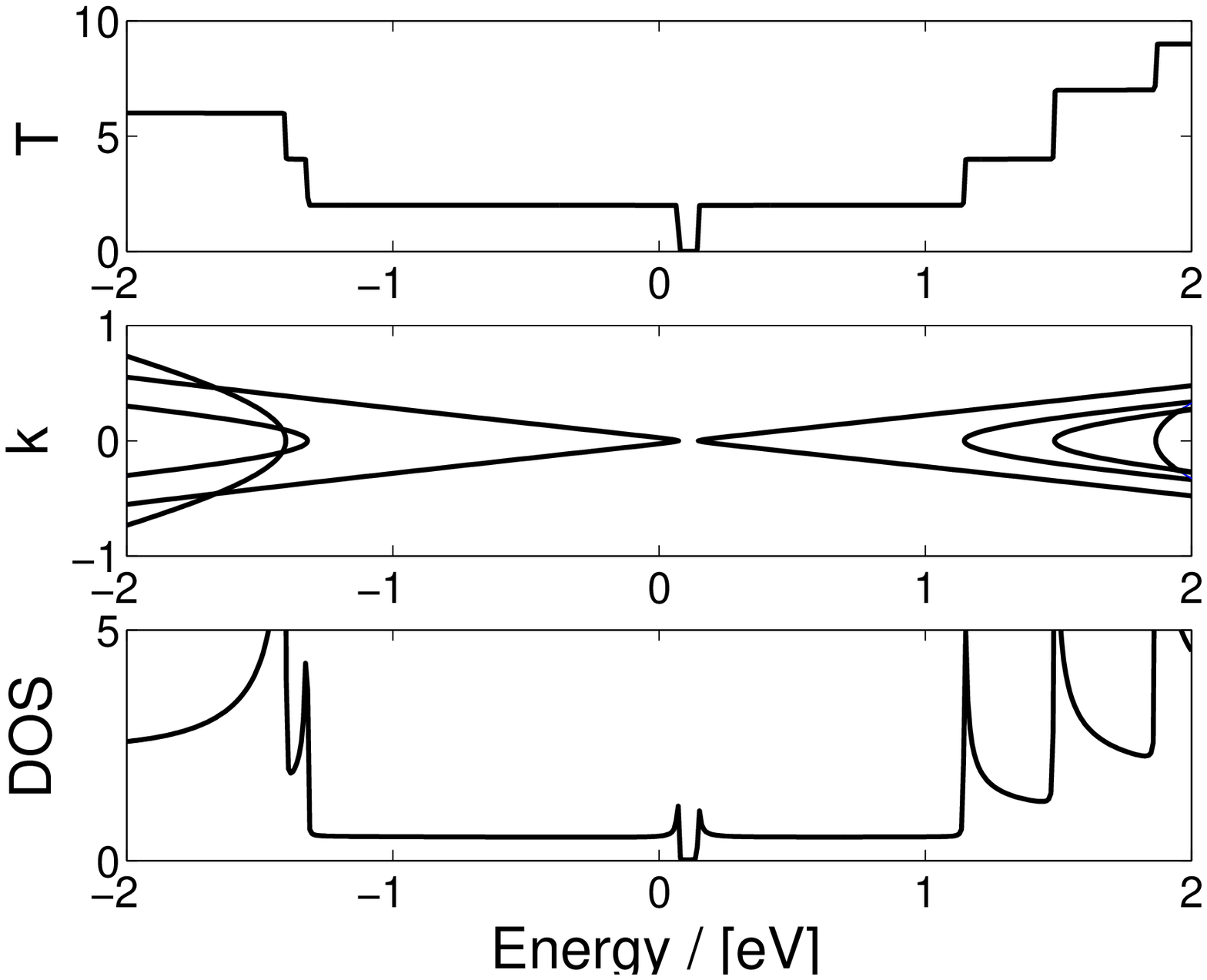}}
\vspace{0.1cm}
\centerline{\epsfxsize=2.6in \hspace{0.5cm}\epsffile{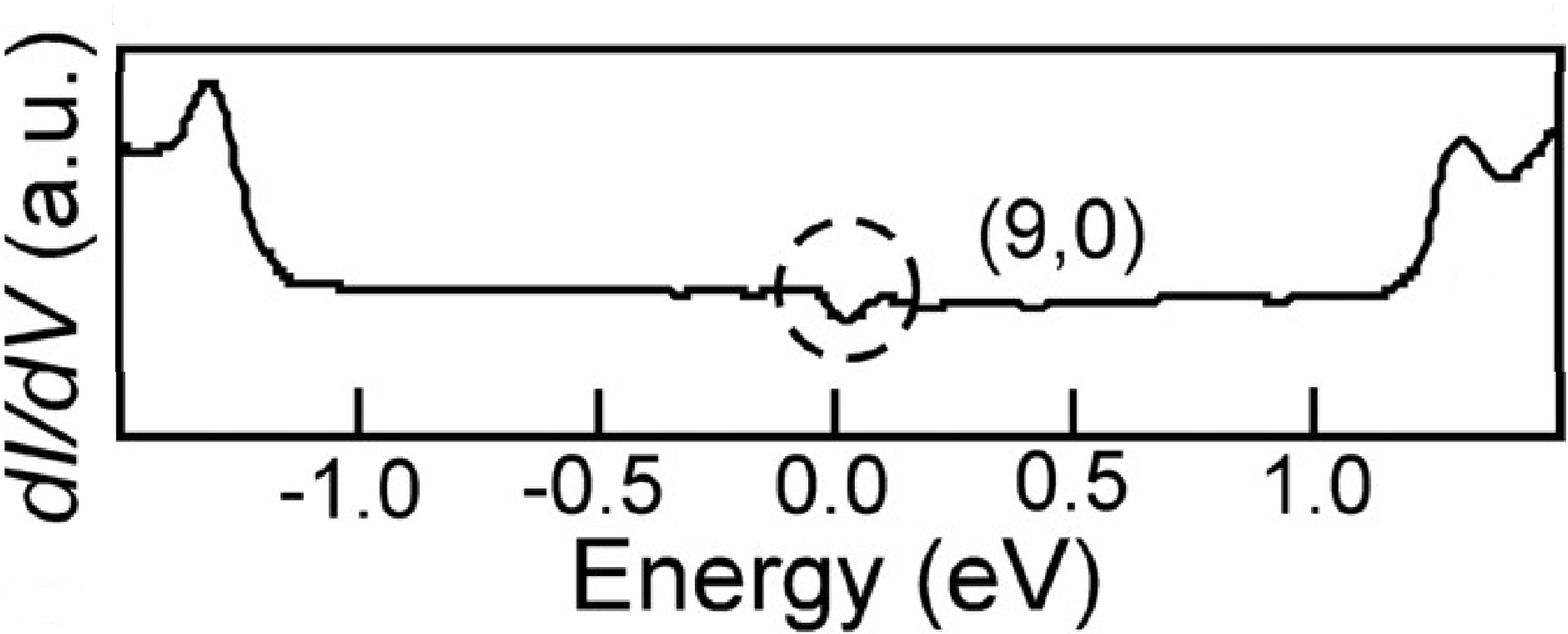}}
\vspace{-0.3cm}
\caption{Bandstructure, DOS, and transmission per spin for a ``metallic'' 
zig-zag $(9,0)$ tube using $sp^3$-orbitals. The gap close to the Fermi level at $E_F=0.0$ eV 
is about $80$meV similar to STS-$dI/dV$ measurements as shown at the bottom\cite{LieberSTS}.
The bottom figure has been ``Reprinted (abstracted/excerpted) with permission from 
M. Ouyang, J.L. Huang, C.L. Cheung, and C.M. Lieber, ``Energy Gaps in Metallic 
Single-Walled Carbon Nanotubes,'' {\em Science}, {\bf 292}, 702 (2001). 
Copyright 2001 AASS.''} 
\label{Fig_N9M0}
\end{figure}

The more complex $sp^3$- and the $sp^3 d^5$-EHT models we are using naturally 
account for these structural deformations (Figs.~\ref{Fig_N12M0} and \ref{Fig_N9M0})
through the structure-dependent overlap matrix ${\bf S}$, cf. Eq.(\ref{EHT_Rule}), yielding 
for our bulk graphene parameters a gap that actually compares quite well quantitatively 
in the case of a $sp^3$-model with experimental scanning tunneling spectra 
(cf. Table \ref{TableGaps}), whereas our $sp^3 d^5$-model shows a poorer quantitative 
match for these class of tubes.
%

\subsection{Ultrasmall Diameter Tubes} 

\begin{figure}[htbp]
\vspace{-0.5cm}
\centerline{\epsfysize=2.75in
\epsffile{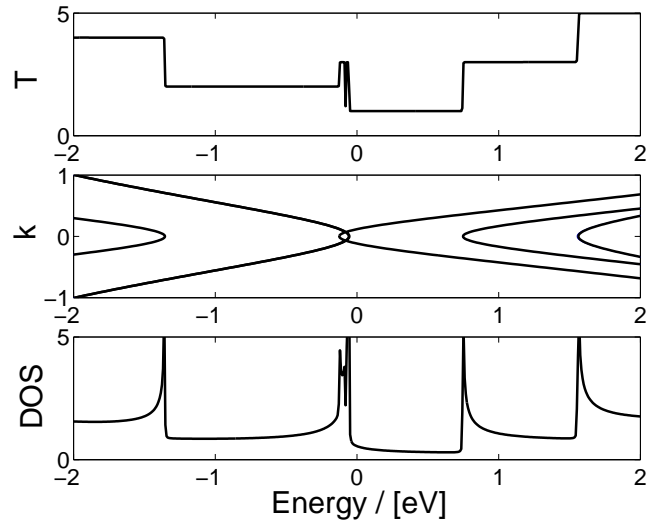}}
\vspace{-0.2cm}
\caption{Bandstructure, DOS, and transmission per spin for a zig-zag $(5,0)$
tube using $sp^3$-orbitals. Due to the crossing of the bands around 
$E_F\approx 0.0$ eV the tube becomes metallic.} 
\label{Fig_N5M0}
\end{figure}

\begin{figure}[htbp]
\vspace{-0.3cm}
\centerline{\epsfysize=2.75in 
\epsffile{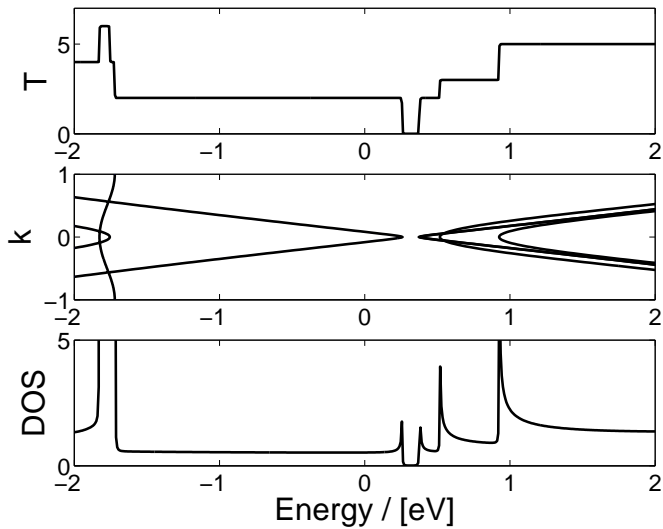}}
\vspace{-0.2cm}
\caption{Bandstructure, DOS, and transmission per spin for a zig-zag
$(6,0)$ tube using $sp^3$-orbitals. The ``metallic'' tube becomes semi-conducting 
with a bandgap of $\approx 0.1$ eV, and shift of the Fermi-level from $0.0$ eV to 
$\approx 0.3$ eV.}
\label{Fig_N6M0}
\end{figure}

For ultrasmall nanotube diameters, hybridization effects start to
become dominant. Fig. \ref{Fig_N5M0} shows that the $(5,0)$ zig-zag
tube, semi-conducting in a zone-folding method, is predicted by EHT to
become metallic for the case of a sp-orbital model, since the valence and 
conduction bands cross at $E_F \approx 0.0$ eV. On the other hand, the zig-zag 
$(6,0)$ tube moves the other way (Fig.~\ref{Fig_N6M0}), from being metallic in a 
zone-folding method to semi-conducting in EHT with a bandgap of $\approx 0.12$ eV
due to strong hybridization effects. DFT-GGA calculations, however, show that a $(6,0)$-tube remains 
metallic due to re-hybridization.\cite{BlaseHybrid,SunGGA} Using in turn spd-orbitals for carbon,
our EHT-model agrees with the DFT-GGA results for these small diameter tubes.
While the accuracy of DFT for semiconducting bandgaps is itself open to question, the contradictory
result for the sp-orbital model could also imply that the inclusion of deformation effects
through just the off-diagonal EHT Hamiltonian elements is no longer a
valid assumption, and the onsite energies themselves need to be
recalculated self-consistently to include the effect of the electrostatics
on the local atomic potentials.
%
%

%
%

%
\begin{table*}[htbp]
\begin{tabular}[t]{ c | c | c | c | c | c | c }
\hline\hline
$(n,m)$ & TB & CNT-bands & DFT & Experiments & EHT-sp & EHT-spd \\
\hline \hline
$(5,0)$  &       $-$                     & $1.91~(\gamma=2.5)$ & $0.0$\cite{SunGGA}             
& $-$                     & $-0.05$ & $0.0$    \\
         &                               & $2.29~(\gamma=3.0)$ &                                
&                         &         &          \\  \hline
$(6,0)$  & $0.05$\cite{BlaseHybrid}      & $0.0$               & $-0.83$\cite{BlaseHybrid}      
& $-$                     & $0.12$   & $0.0$    \\
         & $\approx 0.2$\cite{HamadaTB}  &                     & $0.0$\cite{SunGGA}             
&                         &              &     \\
         & $0.179$\cite{CaoSP3S}         &                     &                                
&                         &              &     \\ \hline
$(9,0)$  & $0.2$\cite{SunGGA}            & $0.0$               & $0.2$ (GGA)\cite{SunGGA}       
& $0.080 \pm0.005$\cite{LieberSTS}  & $0.075$ & $0.13$   \\
         & $0.075$\cite{CaoSP3S}         &                     & $0.17$ (LDA)\cite{BlaseHybrid} 
&                         &              &     \\
         & $0.07$\cite{BlaseHybrid}      &                     & $0.12$ (LDA)\cite{SaitoGW}      
&                         &              &     \\
         & $\approx 0.04$\cite{HamadaTB} &                     & $0.17$ (GW)\cite{SaitoGW}       
&                         &              &     \\ \hline
$(10,0)$ & $0.65$\cite{YorikawaTB}       & $0.91~(\gamma=2.5)$ & $0.88$ (GGA)\cite{SunGGA}      
& $0.83$\cite{LieberSTS2}  & $0.91$  & $0.95$   \\
         & $0.87$\cite{YorikawaTB}       & $1.09~(\gamma=3.0)$ & $0.8$ (GGA)\cite{AvramovGGA}   
&                         &              &     \\
         & $0.85$\cite{HamadaTB}         &                     &                                
&                         &              &     \\ \hline
$(12,0)$ & $0.08$\cite{SunGGA}           & $0.0$               & $0.08$ (GGA)\cite{SunGGA}      
& $0.042 \pm 0.004$\cite{LieberSTS} & $0.045$ & $0.077$  \\
         & $0.0$\cite{HamadaTB}          &                     & $0.057$ (LDA)\cite{MiyakeLDA}  
&                         &              &     \\ \hline
$(13,0)$ & $\approx 0.7$\cite{HamadaTB}  & $0.70~(\gamma=2.5)$ & $0.73$ (GGA)\cite{SunGGA}      
& $-$                     & $0.71$  & $0.74$   \\
         &                               & $0.84~(\gamma=3.0)$ &                                
&                         &         &          \\ \hline
$(15,0)$ & $0.0$\cite{HamadaTB}          & $0.0$               & $0.14$ (GGA)\cite{SunGGA}      
& $0.029 \pm 0.004$\cite{LieberSTS} & $0.026$ & $0.05$   \\
         &                               &                     & $0.038$ (LDA)\cite{MiyakeLDA}  
&                         &              &     \\ \hline
$(16,0)$ & $-$                           & $0.56~(\gamma=2.5)$ & $0.61$ (GGA)\cite{SunGGA}      
& $0.65 \pm 0.30$\cite{DekkerSTS}  & $0.59$  & $0.6$    \\
         &                               & $0.68~(\gamma=3.0)$ &                                
&                         &         &          \\ 
\hline\hline
\end{tabular}
\caption{Comparison of experimentally and theoretically determined
bandgaps (in units $[eV]$) using different theoretical methods:
TB denotes orthogonal tight-binding, CNT-bands refers to a simple 
$\pi$-orbital description with one hopping parameter $\gamma$, DFT 
is density-functional theory using different approximations for 
exchange-correlation potential, and EHT with sp- or spd-orbitals for 
carbon.\cite{CerdaWEB}}
\label{TableGaps}
\end{table*}

For large deformations a fully self-consistent calculation of the
bandstructure might be necessary to describe the electronic structure
properly. If the tube has a large curvature, the respective charge
distributions inside and outside the tube become asymmetric
\cite{BlaseHybrid} leading to the formation of dipoles. Due to the
charge re-distribution and the dipolar electric fields the individual
bands are shifted such that the $(6,0)$ tube remains metallic in
DFT-GGA. The processes of charge-redistribution, dipole-formation, and
the floating of the bands, however, require a fully self-consistent
bandstructure calculation that can be set up within the Poisson NEGF
approach,\cite{DamleDFT} but has not yet been incorporated within our
present computational scheme.
Note, that this is not a shortcoming of the Extended H\"uckel Theory
itself, but arises instead from the importance of self-consistent
electrostatic effects that have been discarded for simplicity in the
bandstructure calculation. Table \ref{TableGaps} compares the bandgaps
for the studied tubes both with respect to STS experiments as well as
other theoretical calculations. Our results based on Extended-H\"uckel
Theory are in good agreement with more sophisticated DFT approaches and
experiments -- in fact, local density approximation (LDA) suffers from
the well-known bandgap problem that an EHT approach calibrated to
graphene seems to bypass. This suggests that an EHT scheme,
supplemented by self-consistent electrostatics, might be a good
compromise between simple $\pi$-orbital tight-binding theories and
computationally expensive DFT-methods, the practicality of the scheme
being particularly important when modeling transport through large
complex nanostructures.

\section{Nanotubes as chemical sensors}

\begin{figure}[htbp]
\centerline{
\epsfysize=1.4in \epsffile{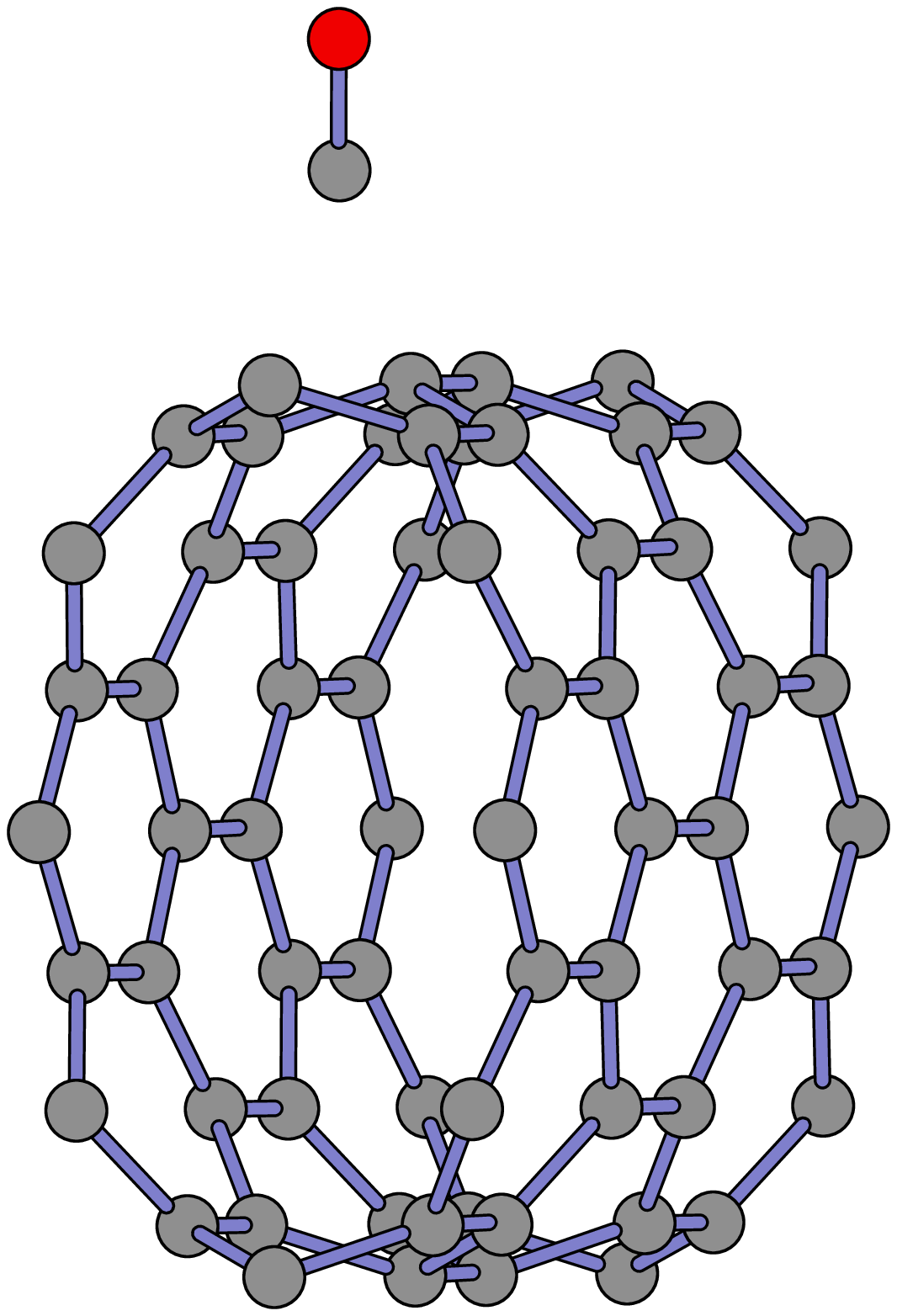}
\hspace{1cm}
\epsfysize=1.0in \epsffile{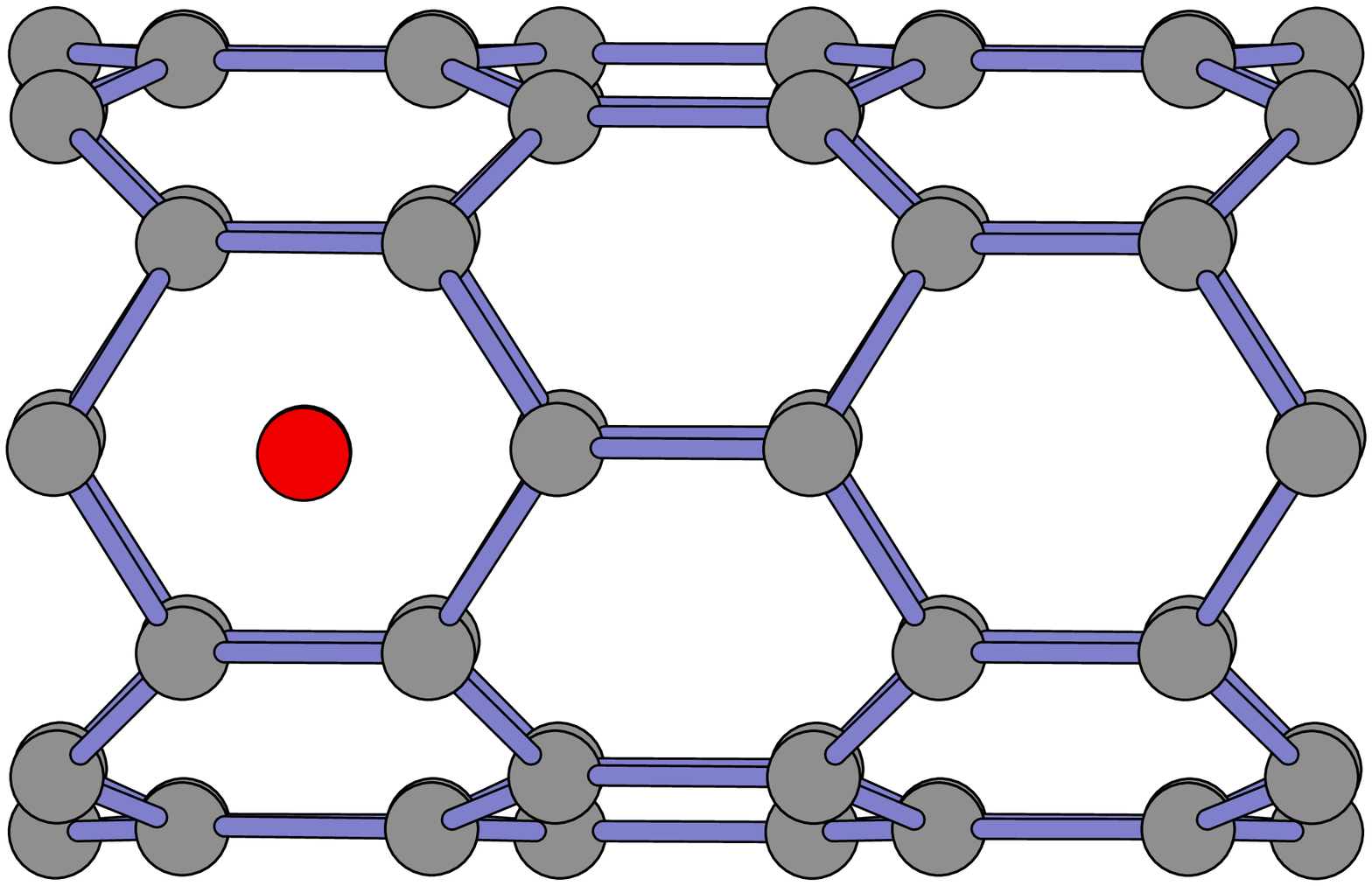}}
\caption{Unit cell of a semi-conducting $(8,0)$ CNT with one CO molecule at 
distance of $d=1.85$ \AA~ from the CNT surface. The CO molecule has been placed above 
the center of the hexagon which is the most stable equilibrium position after
relaxation\cite{DaSilvaCO}. The unit cell of the periodic structure consist of 
two $(8,0)$ CNT unitcells and one CO molecule, i.e. the cell contains $66$ atoms.} 
\label{Fig_SketchCNTMolecule}
\end{figure}

While the previous examples test our EHT prescription for bandstructure
calculations of bare CNTs, we now combine it with molecule chemistry.
It has been suggested that the chemisorption of CO molecules could be enhanced 
by deforming the CNT, so that the regions of highest curvature have an enhanced 
chemical reactivity.\cite{DaSilvaEnergetics} Based on a first principles calculation 
within GGA, da Silva {\it{et al.}} have shown that a semiconducting $(8,0)$ CNT 
can become metallic upon lateral deformation to an elliptical shape, but thereafter 
the semi-conducting state can be recovered by attaching a CO molecule at the spot 
of highest curvature.\cite{DaSilvaCO} This seems to indicate that highly deformed 
CNTs are possible candidates for sensing CO molecules by means of a bandgap variation 
and correspndong metal-insulator transition. We use the DFT calculations by 
da Silva {\it{et al.}} as a qualitative benchmark to explore the accuracy of EHT for 
electronic structure in the presence of a periodic array of CO molecules, of which one 
unit cell is shown in Fig.~\ref{Fig_SketchCNTMolecule}.

Our starting point is the bandstructure of an undeformed semiconducting
$(8,0)$ tube using spd-EHT parameters optimized for 2D-graphene with
the Fermi-level set at $E_F = -13$ eV.\cite{CerdaWEB}
The transferability of the sp- as well as the spd-orbital EHT-parameters have been
discussed in the previous section. 
Setting our vacuum level as the zero energy reference, we shift the CNT Fermi energy 
by $V_c = +8.5$ eV towards the experimental Fermi-level of 2D-graphene, i.e. $E_F = -4.5$ eV,
employing the modified H\"uckel prescription, cf. Eq.~(\ref{EHT_RuleModified}). 
Following Ref.\cite{DaSilvaCO} we deform the tube perpendicular to its axis so that the 
minor axis is $30\%$ of the original tube radius of $R_t=6.3$ \AA, cf.
Fig.~\ref{Fig_SketchCNTMolecule}. 
%

To ensure that our strongly deformed
structure really corresponds to the local minimum of the total energy,
we optimize the deformed tube unit cell by means of Gaussian 03
\cite{Gaussian03} using the spin unrestricted LDA approach within the
Vosko-Wilk and Nusair (SVWN) approximation. During the optimization we
froze two rows atoms along the opposite extremes of the minor axis,
while edge effects were eliminated from the optimized structure by
employing periodic boundary conditions within Gaussian 03 for structure
optimization, with a translation vector of $4.32$ \AA~along the tube
axis. We note that compared to da Silva {\it{et. al}}\cite{DaSilvaCO}
our system being optimized consists of only one unit cell of a $(8,0)$
tube instead of two.

Since the EHT levels are correct upto an overall shift, we start by
aligning the levels of the CO molecule and the CNT. The highest
occupied molecular orbital (HOMO) of CO, calculated within Gaussian 03
using the Becke 3 parameter exchange with Perdew-Wang 91 correlation
(B3PW91) with 6-31g(d) basis sets is $E_{H} = -10.14$ eV relative to
vacuum. The corresponding value in our EHT parametrization is
$E_{H}=-14.09$ eV, so that we need an overall shift $V_{c,CO} = + 3.95$ eV.
After this shift is executed, the CO molecule is attached to the tube
at the point where its curvature is highest. According to
Ref.~\cite{DaSilvaCO} the most favorable location to place the CO
molecule is above the center of a hexagon, as shown in
Fig.~\ref{Fig_SketchCNTMolecule}. Since we are only interested in the
effect of one single CO molecule on the tube dispersion, we need to
avoid any overlap between neighboring CO molecules.  This is
accomplished by using two CNT $(8,0)$ unit cells and attaching only one
CO molecule. The effective periodicity of the system is then $8.64$ \AA, 
large enough that the neighboring CO molecular basis functions do not overlap.

\begin{figure}[htbp]
\centerline{ \epsfysize=2.6in \epsffile{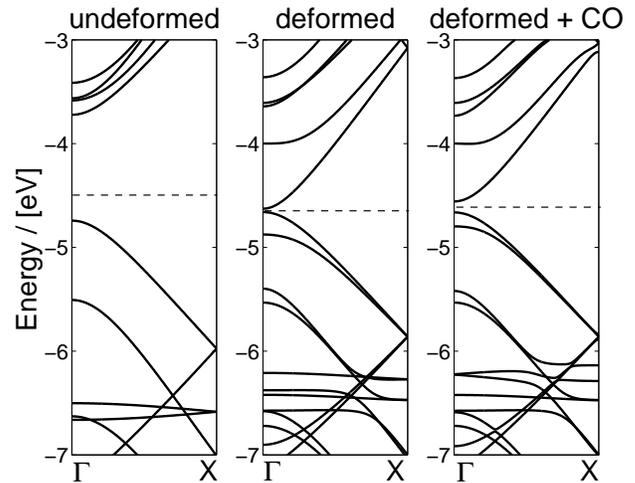}}
\vspace{-0.2cm}
\caption{Bandstructures of a zig-zag $(8,0)$ CNT using $sp^3 d^5$-orbitals. 
Left: undeformed tube with a bandgap of $E_G \approx 1.1$ eV, middle: deformed tube with 
a small gap of $E_G \approx 25$ meV, and right: deformed tube with attached CO molecule 
at distance $1.85$ \AA. The gap here is $E_G \approx 100$ meV. The Fermi energy is indicated 
by the dashed line.} 
\label{Fig_EKCO}
\end{figure}

\begin{figure}[htbp]
\centerline{ \epsfysize=7.2cm \epsffile{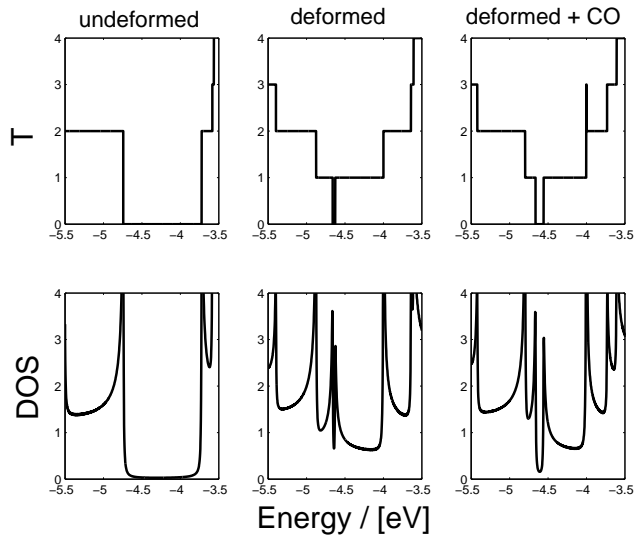}}
\vspace{-0.2cm}
\caption{DOS and transmission per spin for the semi-conducting $(8,0)$ CNT:
left: undefomed, middle: deformed, and right: deformed with attached CO-molecule.} 
\label{Fig_DOSTECO}
\end{figure}

Figure \ref{Fig_EKCO} shows the dispersion, and Fig.~\ref{Fig_DOSTECO} the respective
DOS and transmission per spin for the combined CNT-molecular
system. The original bandgap $E_G=1.1$ eV of the undeformed tube (left)
shrinks down to $E_G \approx 30$ meV upon the $30\%$ lateral
deformation, so that the tube becomes effectively metallic at room
temperature (center). Attachment of the CO molecule on the deformed
CNT (right) makes the tube semiconducting once again, with a bandgap of
$E_G \approx 100$ meV, much less than that of the undeformed tube, but
noticably larger than thermal energies at room temperature. Our results 
agree qualitatively with da Silva {\it{et. al}},\cite{DaSilvaCO,DaSilvaEnergetics}
even though there are quantitative differences: the initial bandgap of the undeformed 
tube is $0.66$ eV and becomes completely closed after lateral deformation, so that the 
tube becomes truly metallic. The recovered bandgap upon CO-attachment
is with $\approx 200$ meV, of similar order as in our case with $100$ meV. 
The largest source of disagreement is the bandgap of the undeformed CNT;
given that DFT bandgaps (see Table~\ref{TableGaps}) are often questionable as well, a
quantitative resolution of this discrepancy may need experimental attention.
%
%

%

Note that the present calculations are all non-self-consistent,
ignoring effects due to the rearrangement of charges during tube
deformation, as well as those arising from charge transfer between the
molecular species and the nanotube, while a proper self-consistent calculation 
as discussed in the last section, is needed to do quantitative justice to this problem.
Nonetheless, the EHT parameters seem to be quite transferable between
bandstructure as well as surface chemistry, in particular for strongly
deformed structures without the need for any reparametrization. This makes 
EHT a good compromise between accuracy and practicality. In
our follow-up paper, we will demonstrate the applicability of EHT to
modeling silicon, including its transferability between bulk and
multiple surface bandstructures of reconstructed silicon surfaces and
also for nanowires. Applying this aproach has allowed us to quantitatively explain 
and in some cases even predict interesting experimental results involving
molecular conductors on silicon.\cite{RakshitNDR,LiangC60}

\section{Future work}

For transport calculations, we often need a minimal model that can do
justice to bandstructure, electrostatic as well as bonding chemistry. 
This becomes particularly important if one wants to deal with strongly 
deformed structures, interfaces or combinations thereof including molecules.
We have shown that Extended H\"uckel Theory provides a good practical
compromise to capture various aspects of bandstructure and chemistry.
The two attributes that make this semi-empirical approach especially
useful are the presence of explicit basis-sets and non-orthogonality.
To make this chemists' approach to bandstructure of further use, it may be
preferable to work with a self-consistent {\it{complete neglect of
differential overlap}} (CNDO) approach to bring in differential Coulomb
costs into the picture.

For many nanoscale structures such as nanotubes and nanowires
perhaps even interfacing with smaller molecules, we believe that a
semi-empirical approach combining bandstructure and chemistry is
essential, given that typical tight-binding theories are not
transferable beyond small deformations while DFT theories are
computationally quite prohibitive. The latter becomes even more
difficult to implement when we want to move from equilibrium electronic 
structure to more complicated nonequilibrium transport problems. 
It is generally acknowledged that much of the conducting properties of 
nanostructures are dominated by their interfaces and contacts. 
It is in this complicated domain that we believe the real strength of a
non-orthogonal theory with explicit basis sets such as EHT or
CNDO is likely to manifest itself.

\section*{Acknowledgement}
We acknowledge the support of the Army Research Office through the Defense 
University Research Initiative in Nanotechnology (DURINT) program, and the 
Army Research Office.
J. Cerda acknowledges support from the Spanish DGICyT and CAM under 
contract No. MAT2004-05348-C04-2 and MAT/0440/2004, respectively.
The authors are indepted to M.P. Anantram for his help and discussions.
We thank K.H. Bevan, H. Raza, L. Siddiqui, and M.S. Lundstrom for helpful 
discussions.



\begin{thebibliography}{99}
\bibitem{LieberSTS2}
T.W. Odom, J.L. Huang, P. Kim, and C.M. Lieber,
``Structure and Electronic Properties of Carbon Nanotubes,''
{\em J.Phys.Chem.B}, {\bf 104}, 2794 (2000).\\
Note: Since the value for the $(8,0)$ CNT is not given explicitly, we have determined 
its bandgap from the STS-plot by measuring the distance between the 1st VHS where the 
STS is neglible, so that a gap can be clearly identified. The distance between the largely 
broadenend peaks is about $1.1$ eV.
\end{thebibliography}
\end{document}